%% file: letterpks.tex
\documentclass[usenatbib]{mn2e}
\usepackage{amsfonts,amssymb,amsmath}
\usepackage{graphicx}

\volume{VER24}

\title[Hydrodynamical model of PKS~1510-089 light curve]
{A hydrodynamical model for the FERMI-LAT $\gamma$-ray light curve of Blazar 
PKS~1510-089}

\author[J. I. Cabrera et al]{J. I. Cabrera$^{1}$\footnotemark[1], 
Y. Coronado$^{1}$\footnotemark[1], E.
Ben\'itez$^{1}$\footnotemark[1], S. Mendoza$^{1}$\footnotemark[1], 
D. Hiriart$^{2}$\footnotemark[1] and M.
Sorcia$^{1}$\thanks{E-mail address: jcabrera@ciencias.unam.mx (JIC),
coronado@astro.unam.mx (YC), erika@astro.unam.mx (EB), sergio@astro.unam.mx
(SM), hiriart@astrosen.unam.mx (DH), msorcia@astro.unam.mx (MS).} \\
$^{1}$Instituto de Astronom\'ia, Universidad Nacional Aut\'onoma de
M\'exico, AP 70-264, 04510 Distrito Federal, M\'exico \\
$^{2}$Instituto de Astronom\'ia, Universidad Nacional Aut\'onoma de M\'exico, 
AP 877, 22800 Ensenada, B.C., M\'exico }

\begin{document}


\pagerange{\pageref{firstpage}--\pageref{lastpage}} \pubyear{2012}

\maketitle

\label{firstpage}

\begin{abstract}
A physical description of the formation and propagation of working
surfaces inside the relativistic jet of  the Blazar PKS~1510-089 are used to
model its \(\gamma\)-ray variability light curve using \textit{FERMI-LAT}
data from 2008 to 2012. The physical model is based on conservation
laws of mass and momentum at the working surface as explained by
\citet{mendoza09}. The hydrodynamical description of a working surface
is parametrised by the initial velocity and mass injection rate at the base of 
the jet. We show that periodic variations on the injected velocity
profiles are able to account for the observed luminosity, fixing model
parameters such as mass ejection rates of the central engine
injected at the base of the jet, oscillation frequencies of the flow and
maximum Lorentz factors of the bulk flow during a particular burst.
\end{abstract}


\section{Introduction}
\label{introduction}

Among all types of AGN, Blazars \citep[Blazar class is defined
as radio loud sources conformed by the BL~Lac objects and the 
Flat Spectrum Radio Quasars -FSRQ, see e.g.][and references
therein]{Fossati97,Ghisellini98} represent the most energetic class. 
They are known to have the most powerful jets \citep[e.g.][]{lister09}
and also show a highly variable Spectral Energy Distribution (SED) 
from the radio to the \(\gamma\)-rays wavelengths 
\citep[see][and references therein]{SED01, dammando11}.

  The FSRQ PKS 1510-089 is known to be one of the most powerful
astrophysical objects with a highly collimated relativistic jet that
has shown apparent superluminal velocities between \( 20 c \) to  \(
46 c \) and with a  semi-angle aperture for the jet \( \sim 0.2^\circ
\) \citep{jorstad05}.  Since the angle between the relativistic jet and
the observer's line of sight \( \sim 1.4^\circ \) -- \( 3^\circ \), the
jet almost coincides with the observer's line of sight \citep{homan02,
marscher10}.  PKS~1510-089 was one of the \( \gamma \)-ray sources
detected by EGRET \citep{hartman99}. It has been monitored at high
energies with \textit{AGILE} \citep{pucella08, dammando08,Lucarelli12} and
by \textit{FERMI-LAT} and \textit{AGILE} \citep{tramacere08, ciprini09,
dammando09}. It has also been studied with \textit{MAGIC} and HESS
\citep{cortina12, Wagner10}. The most prominent outbursts displayed
by PKS 1510-089 were reported by \citet{Kataoka08}, \citet{ciprini09}
and \citet{orienti12}.  The high activity observed in this source,
turns it into an ideal target for the physical study of its highly
relativistic jet.

Precise models for the light curve (LC) produced by the outburst and
flares from Blazars are not done using directly the data variations
observed in different wavelengths. Instead, models are applied to
explain the behaviour of the SED \citep[e.g.][]{SED01,dammando11}.
Direct understanding of the LC requires a precise knowledge of the
hydrodynamical behaviour of the relativistic flow.  \citet{mendoza09},
hereafter M09, have constructed a hydrodynamical model of the motion
of a working surface inside a relativistic jet  which is able to fit
the observed LCs of long Gamma-Ray Bursts (lGRB's).  Since the jets in
Blazars are highly relativistic and their jet is nearly pointing towards
the observer, similar to the jets observed in lGRB's, the physical
ingredients of both phenomena can be considered the same but occurring
at different physical scales of energy, sizes, masses, accretion rates,
etc. \citep[cf.][]{mirabel02}.

The Blazar PKS~1510-089 is of tremendous importance since
it exhibits extreme relativistic motions.  
As such, its energy curve must present luminosity variations
and periods of extreme activity displayed as outbursts that, when
physically modelled, can yield a better understanding of
the physical parameters associated to the mechanism producing
the observed luminosity.

In this letter, we assume that the mechanism producing the observed LC in 
a typical lGRB is exactly the same that produces the variable LC of the 
Blazar PKS~1510-089. We thus apply the hydrodynamical
jet model presented in M09 to the LC variations
displayed by the Blazar PKS~1510-089 in the $\gamma$-ray domain,
using public data obtained with the \textit{FERMI-LAT} telescope.

The letter is organised as follows. In Section~\ref{fermi-lat-data}
we explain in general terms the data reduction process. In
Section~\ref{model} we describe the characteristics of our hydrodynamic
model. The fit done to the data with the hydrodynamic model is explained
in Section~\ref{sec:Fit}.  The results of our fits and the discussion
of the main physical parameters obtained in the modelling are
presented in Section~\ref{discussion}.  Throughout this letter we use a
standard cosmology with \( H_0 = 71 \, \textrm{km}\, \textrm{s}^{-1} \,
\textrm{Mpc}^{-1}\), \( \Omega_\textrm{m}=0.27 \) and \( \Omega_{\lambda}
= 0.73\) \citep[see e.g][and references therein]{Kataoka08}.

\section[]{Fermi-LAT Data}
\label{fermi-lat-data}

  The gamma-ray fluxes were obtained in the range of 0.2 to 300 GeV
using the public database of \textit{FERMI-LAT} from 2008 August 08
to 2012 May 28.  The data were reduced with the \textit{FERMI} science
tool package \citep[see e.g.][]{Atwood2009} in the same energy range,
taking into account the diffuse galactic background radiation, the
instrument response matrix p7v6, and considering a zenith angle $
< 105^\circ $. We also calculated the active time of the detector
and the PSF. The $\gamma$-ray LC was constructed modelling the
flux with a power law of the form  $\mathrm{d} N / \mathrm{d} E =
N_{0}(E/E_{0})^{\gamma}$, with \( \gamma = 2 - 3 \) in accordance with
the results of \citet{abdo10}.  The fluxes and errors obtained with this
package are given in photons~$\times$~cm$^{-2}$\,s$^{-1}$. For further
physical interpretation of the data, we have converted these fluxes and
errors to MeV cm$^{-2}$\,s$^{-1}$.

\begin{figure*}
\begin{flushleft}
  \includegraphics[scale=0.65]{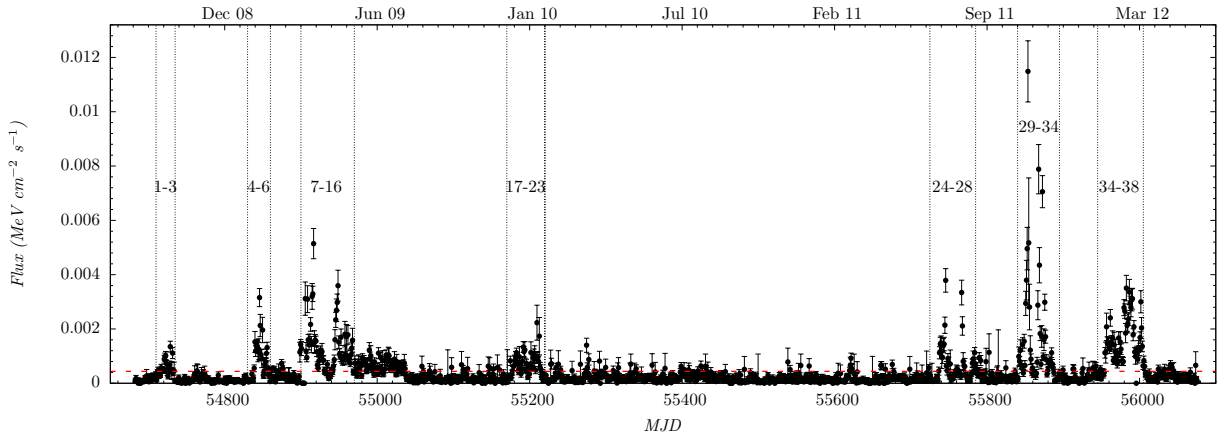} 
\end{flushleft}
\caption{Fermi-LAT light curve of Blazar PKS~1510-089 (from 0.2 to 300 GeV)
obtained from 2008 August to 2012 May. The outburst identification number
(ID\#) labelled in the figure stands for the different flares selected in
our work (see text).  The 3\(\sigma\) noise level is represented by the
red horizontal dashed line. }
\label{Fig0}
 \end{figure*}

 The photons considered for analysis were taken from a region centred
on the coordinates of PKS~1510-089 with a radius of \( 15^\circ \). 
Figure~\ref{Fig0} shows the $\gamma$-ray  LC, with a bin size of 1
day. We chose these bins, since the errors are larger using shorter
bin sizes, complicating the analysis of the data and because particular
outbursts can be adequately resolved.

  From Figure~\ref{Fig0} it follows that the source displayed the
historical maximum outburst in  MJD 55851, corresponding to 2011 October
17 and reported by \citet{hungwe2011}.  Another important outburst
occurred in MJD 54899 (2009 March 9) and was observed with \textit{AGILE}
\citep{dammando09}. Several flares or outbursts can be observed in the
LC. The most relevant events occurred in MJD 54717 (2008 September 8), MJD
54843 (2009 January 12), MJD 55200 \citep[2010 January 4,][]{benitez11},
MJD 55730 (2011 June 18), and MJD 55954 (2012 January 28).  This last
event was also observed by \textit{AGILE} \citep{verrecchia12} and
\textit{MAGIC} \citep{cortina12}.  Note that \citet{marscher10} reports
extra flares \( < 200 \text{MeV} \) during the period 54850 - 54950 MJD,
which are not seen in our \( > 200 \text{MeV}  \)  selection.

%
\section[]{A hydrodynamical model for the Light Curve of PKS~1510-089}
\label{model}
The formation of internal shock waves on a relativistic jet are
commonly explained by different mechanisms, such as the interaction of
the jet with inhomogeneities of the surrounding medium, the bending
of jets and time fluctuations in the parameters of the ejection (see
e.g. \citealt{rees94,mendoza-1-02,jamil08,mendoza09}). In particular,
the model by M09 is a hydrodynamical description that can
be applied to shock waves inside relativistic jets. This semi-analytical model
describes the formation of a working surface inside a hydrodynamical jet
due to periodic variations of the injected flow.  When fast flow overtakes
slow flow, an initial discontinuity is formed and a working surface
(two shock waves separated by a contact discontinuity) is produced.  The
working surface travels along the jet and radiates away kinetic energy.
The article by M09 assumed that the efficiency converting factor is \(
\sim 1 \) and that it is mostly emitted in the \( \gamma \)-ray band.
As explained in Section~\ref{introduction}, the Blazar PKS~1510-089
behaves as an scaled typical lGRB and as such, the hypothesis used by
M09 can be extended to this particular object.  As we will
discuss in section~\ref{discussion}, this assumption is coherent with
the physical properties found from the model.
Following M09, we assume that flow is injected at the base
of the jet with a periodic velocity given by
\begin{equation} 
  v(\tau)= v_0 + c \eta^2 \sin \omega \tau, 
\label{eq01}
\end{equation}

\noindent where \( \tau \) is the time in the rest frame of the source,
the velocity \( v_0 \) is the ``background'' bulk velocity of the flow inside
the jet, and \( \omega \) is the oscillation frequency. The positive
constant parameter \( \eta^2 \) is chosen in such a way that oscillations
of the flow are small so that the bulk velocity \( v(\tau) \) of the flow
does not exceed the velocity of light \( c \).  The mass ejection rate \(
\dot{m}(\tau) \) from the central engine which is injected at the base of
the jet is assumed constant through a particular outburst event, but is
allowed to vary from one outburst to another.  The radiated energy of the
flow as a function of time is calculated as the difference between the
total energy \( E_0 \) injected at the base of the jet and the kinetic
energy inside the working surface \( E_\textrm{ws} \).  The luminosity
\( L \) is thus calculated as the derivative of this radiated energy
with respect to time.   As described by M09, there are two ways of calculating
this luminosity curve.  The first method consisted in a semi-analytical
procedure and the second is performed with a full hydrodynamical numerical
model.  The authors showed that the semi-analytical model is in good
agreement with the full numerical simulation, and as such we model the
LC of PKS~1510-089 using their semi-analytical approach.

  The semi-analytical approach is based on the assumption that
equation~\eqref{eq01} is valid and as such, one needs to know (or find
through fits to observational data) the values of \( v_0 \), \( \eta^2
\), \( \omega \) and \( \dot{m} \).  Furthermore,  the mass ejection
rate \( \dot{m} \) enters in the description of the problem through
the luminosity relation: \( L \propto \dot{m} c^2 \).  The average
bulk velocity \( v_0 \) must come from observational data \cite[for
this particular source][reports a value $\Gamma(v_0)=18$]{dammando08}.
With this, the model is left with three free parameters: \(\eta^2\),
\( \dot{m} \), and \( \omega \), which can be fixed by fitting the best
theoretical LC to the observational data.

\section{Modelling the $\gamma$-ray Light Curve} 
\label{sec:Fit}

 To model the LC of Figure~\ref{Fig0}, we have selected the most
conspicuous flares. The criterion used consists of selecting
only those flares that are beyond $3 \sigma$ noise level
according to the errors shown in the LC. By doing so, it turns out that
38 relevant peaks were chosen for our fitting.

 As explained in section~\ref{model}, the model has four free parameters.
The velocity parameter \( v_0 \) for this particular object is such that
its Lorentz factor is \( \Gamma(v_0) =18 \).  To  calculate the measured
luminosity \( L \) from the observed flux \(F\), we multiply the observed
flux \( F \)  by \citep{dermer01,dermer02,longair,Ghisellini93}: \( 4\pi D_{L}^2
\delta^{-p} \) where the relativistic beaming \( \delta :=  1 /
\Gamma(v_0) \left( 1 - \left( v_0/c\right) \cos \theta \right) \sim 18 \),
for a  luminosity distance $ D_{L} $, which for this particular case is
\(D_{L} = 1919 \, \textrm{Mpc}\) and the angle \( \theta \sim 1.4^\circ -
3^\circ \) is the angle between the jet and the observer's line of sight
(cf.  Section~\ref{introduction}). We have selected a beaming index
\( p = 3 \) in accordance with the results of \citet{Wu11} for Blazars and
lGRB's.

\begin{figure*}

 \begin{tabular}{cc}
 
  \centering

  \includegraphics[scale=0.51]{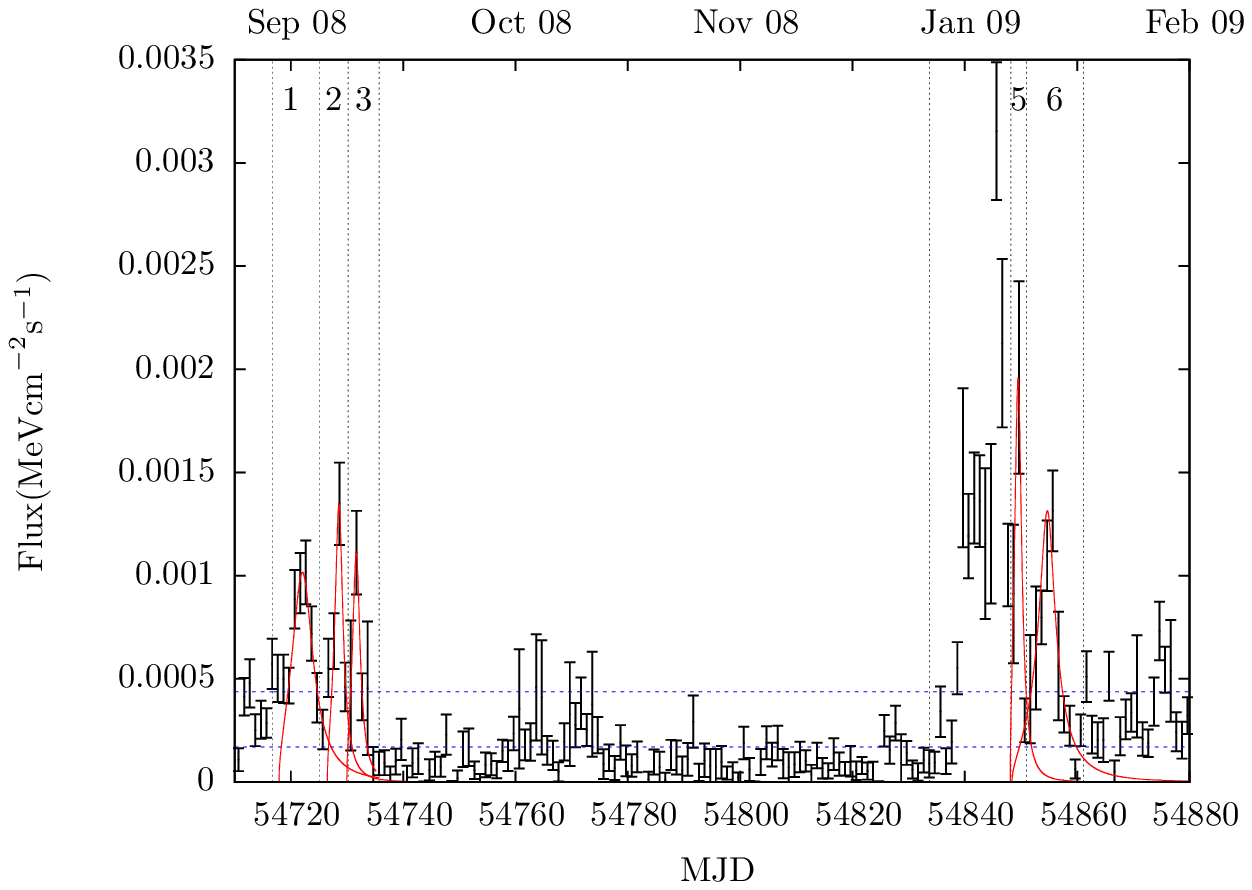} &
  \includegraphics[scale=0.51]{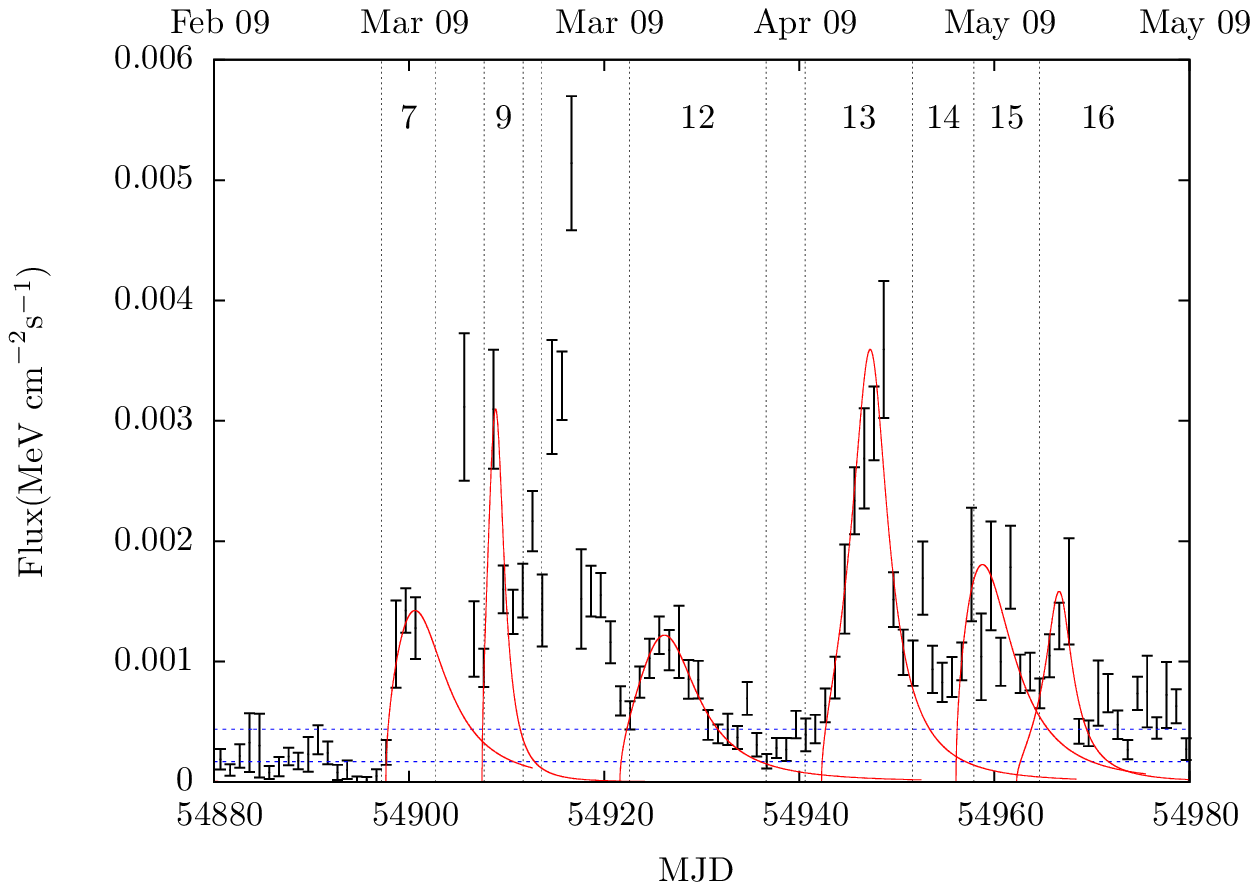} \\
  \includegraphics[scale=0.51]{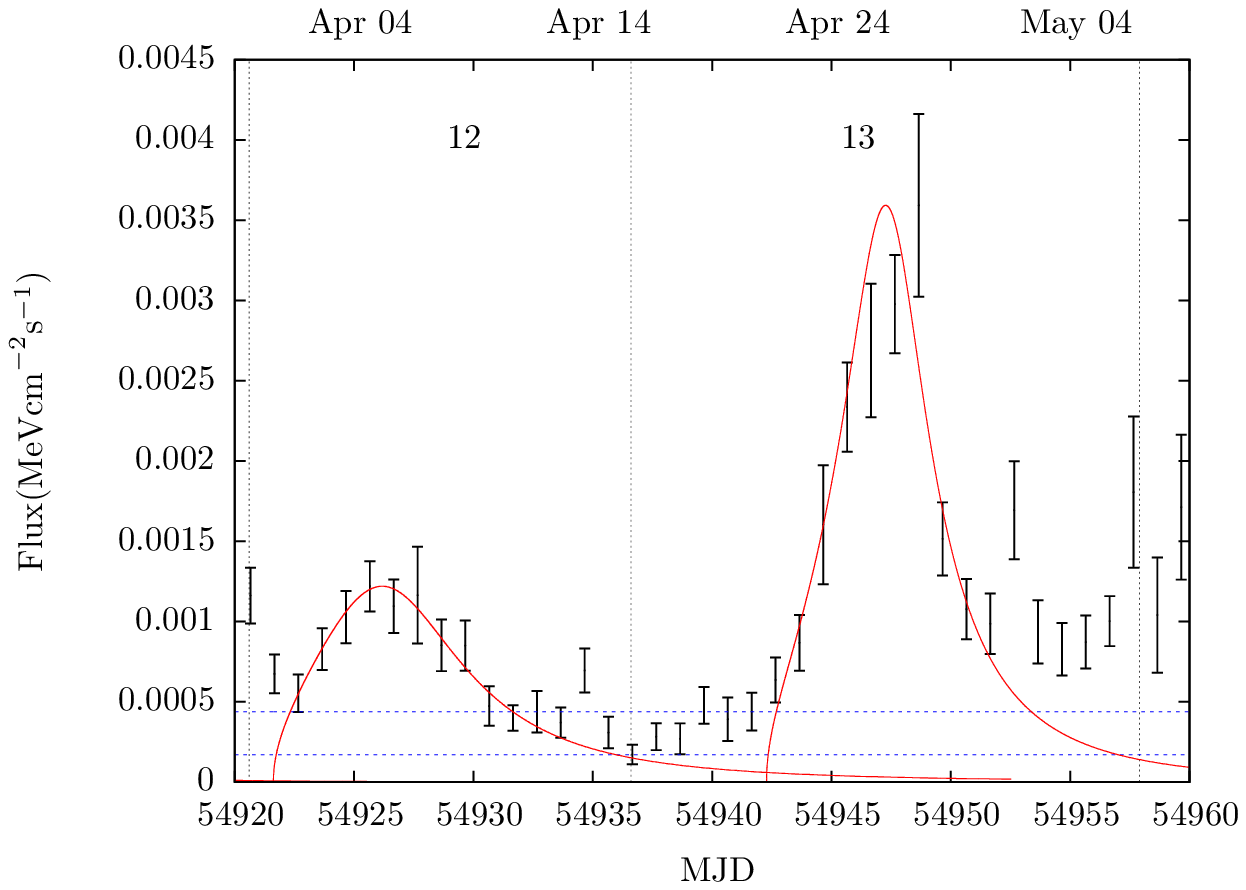} &
  \includegraphics[scale=0.51]{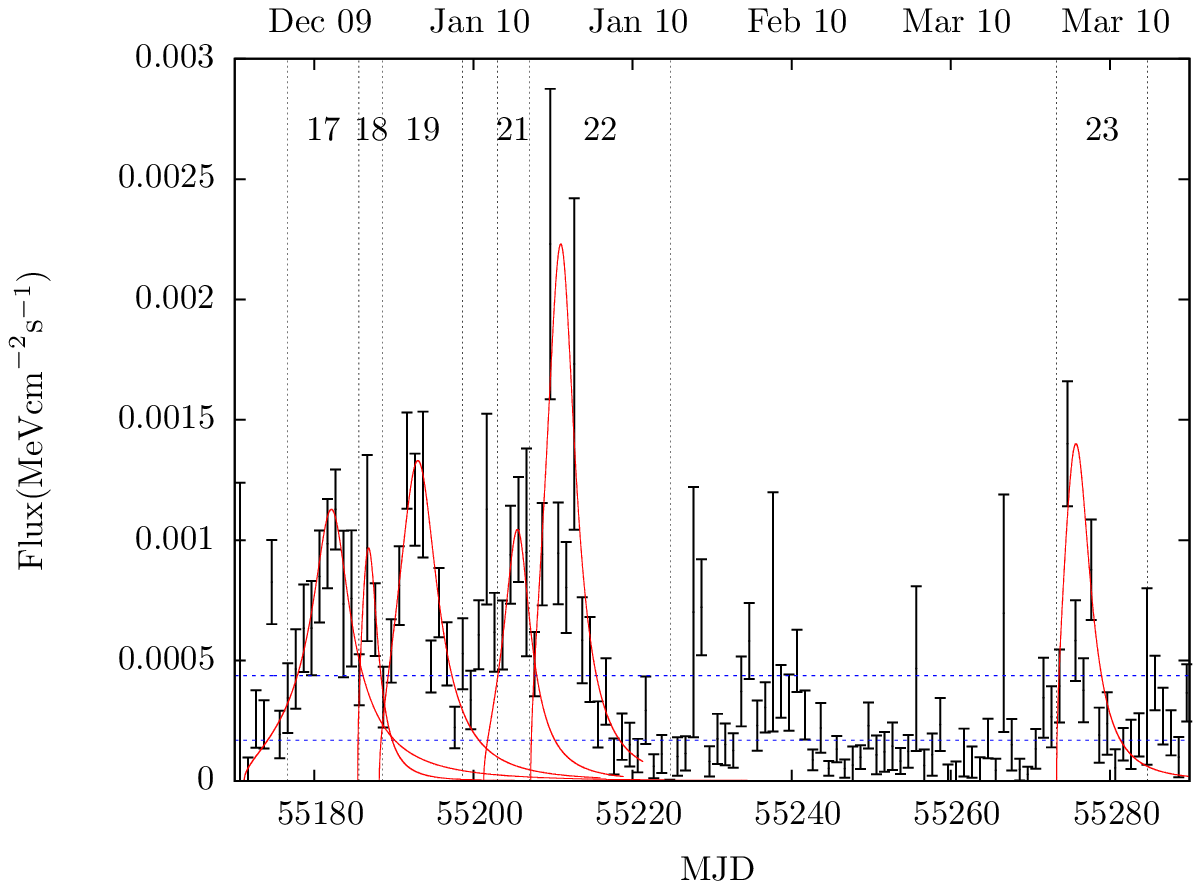} \\
  \includegraphics[scale=0.51]{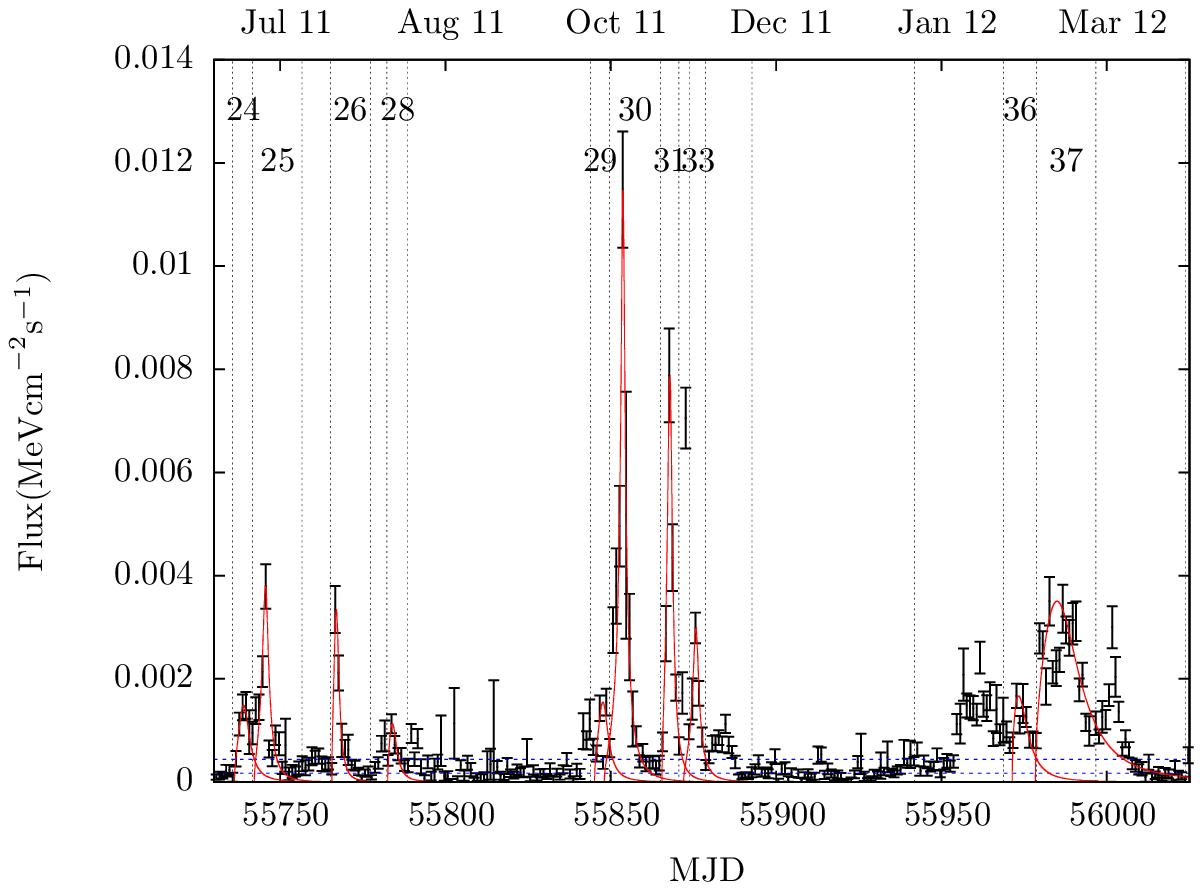} &
  \includegraphics[scale=0.51]{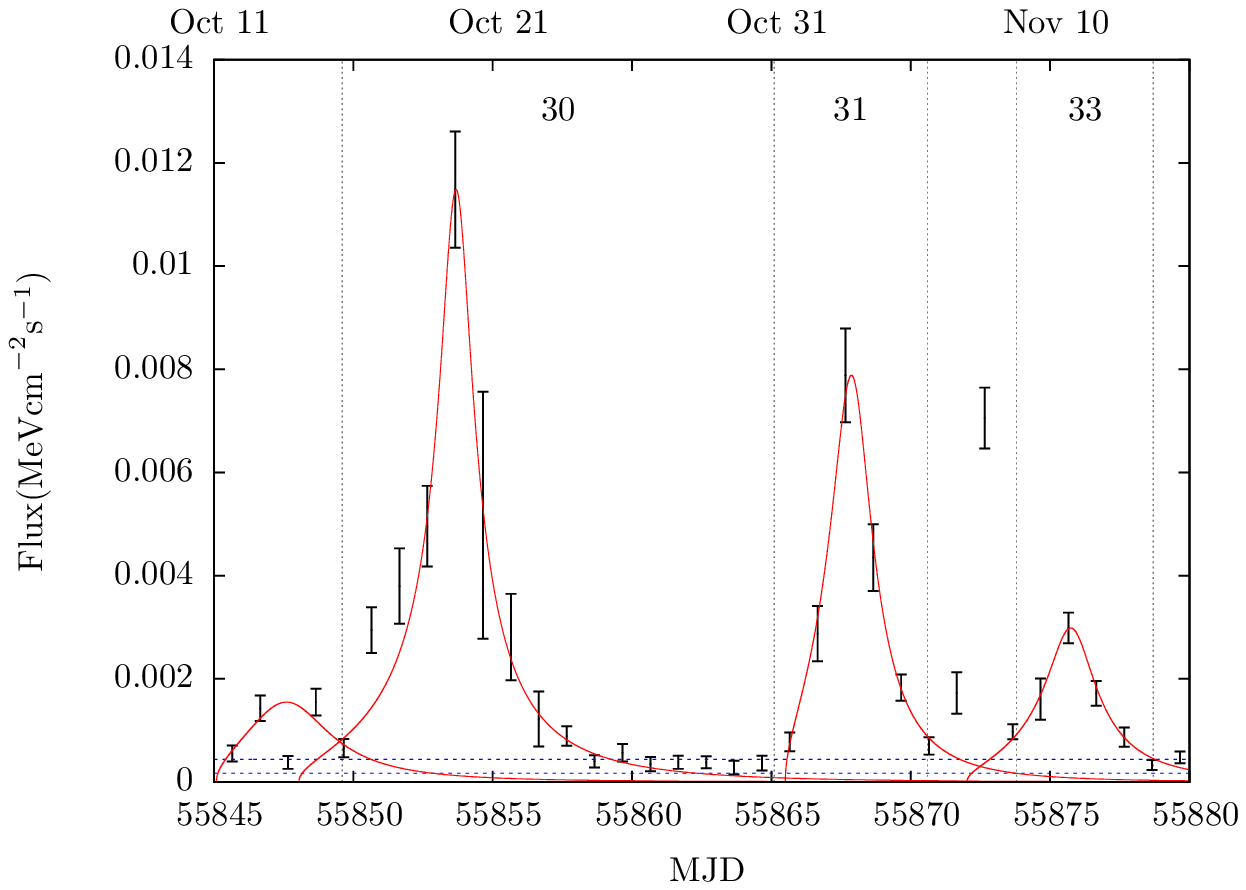}
 \end{tabular}

\caption{In each panel, the continuous red curve represents the fitting
done to the light curve of PKS~1580-089 with the semi-analytical model
of internal shock waves (working surfaces) by \citet{mendoza09}. Blue
horizontal dotted-lines in all panels show the  $1~\sigma$ and  $3~\sigma
$ noise levels. Top-left panel shows variations from 2008 September  to
2009 February.  Top-right shows variations from 2009 February to 2009 May.
The central left-panel shows a zoom of the peaks 12 and 13.  The central
right-panel shows a few outbursts observed from 2009 December to 2010
March. Bottom left-panel shows recent variations occurred from 2011 July
to 2012 March. Finally, bottom-right panel shows a zoom of the prominent
October 2011 outburst. This outburst is $\sim$ three times more luminous
than the one observed in 2009 March.  Up to now, this is the most violent
outburst observed in the $\gamma$-ray  waveband by \it{FERMI}.}

  \label{Fig2}
\end{figure*}

 The model presented by M09 is such that the theoretical luminosity and
time are presented in a very particular system of units.  To fit the best
theoretical LC to the data, one needs to have a common system of units.
To achieve this, we have normalised the ``measured''  Luminosity to its
peak and the measured time to the FWHM of the measured LC.  In order
to compare with the theoretical model, the theoretical LC is also
normalised to its peak and the time is normalised to the FWHM of the
theoretical luminosity curve.  Once both theoretical and measured LCs
are in this common dimensionless system of units, this procedure allows
us to fit the best theoretical LC by performing a \( \chi^2 \)
statistical test to find the optimal parameter \( \eta^2 \).  Note that
in this normalised system of units, the model only depends on one free
parameter: \( \eta \).  Once the value of \( \eta \) is found, we can
rescale back to physical units and in such a rescaling the parameters \(
\dot{m} \) and \( \omega \) are obtained, since according to M09, \( L
\propto \dot{m} c^2 \) and \( t \propto \omega^{-1} \).  The luminosity
fits are then transformed to the observed flux dividing them by \( 4
\pi D_L/ \delta^{3+\alpha} \).  The results of these fits are shown in Figure
\ref{Fig2}.  The obtained values of the physical parameters of the model
for each particular modelled outburst are presented in Table~\ref{table1}.

There are a certain subclass of outbursts that we do not model.  These
outbursts, labelled 8, 10, 20, 27 and 32 in Figure~\ref{Fig0}, do not have
enough data to allow us an accurate modelling.  The outburst labelled 11
seems to have a fall that develops into a constant value before reaching
an expected minimum and no data points further, so it seems incomplete.
Outburst 14 has huge errors and the \( \chi^2 \) statistical test does
not converge.  Outbursts 34 and 35  have large errors which also
makes the modelling not accurate.

\begin{table}

 \begin{center}
 \begin{footnotesize}
 
 \begin{tabular}{@{}lrrcrrc}

  \hline

  Date & ID & MJD  & $ \eta^{2}/c $ & $\Gamma_\text{max}$ &  $\omega^{-1}$& $\dot{m}$  \\ 
  & $\#$ & \text{+54000} & \( 10^{-3} \) &  &$10^{3}\textrm{s}$ & $
  10^{-3} \textrm{M}_{\odot}/\textrm{yr} $\\

  \hline

  08 Sep & 1  & $722.66$   & $1.500$ & $106$ & $1.05$ & $ 2.16  $ \\

  08 Sep & 2  & $728.66$   & $1.520$ & $143$ & $0.50$ & $ 2.87 $ \\

  08 Sep & 3  & $731.66$   & $1.510$ & $120$ & $0.41$ & $ 2.37 $ \\

  09 Jan & 5  & $849.66$   & $1.501$ & $107$ & $0.34$ & $ 4.18 $ \\

  09 Jan & 6  & $855.66$   & $1.533$ & $209$ & $1.49$ & $ 2.80 $ \\                  

  09 Mar & 7  & $899.66$   & $1.330$ & $48$  & $0.94$ & $ 3.04 $ \\

  09 Mar & 9  & $908.66$   & $1.460$ & $76$  & $0.37$ & $ 6.61 $ \\                           

  09 Apr & 12 & $925.66$   & $1.430$ & $66$  & $1.27$ & $ 2.60 $ \\

  09 Apr & 13 & $948.66$   & $1.515$ & $130$ & $1.22$ & $ 7.67 $ \\

  09 May & 15 & $957.66$   & $1.300$ & $45$  & $0.88$ & $ 3.85 $ \\

  09 May & 16 & $967.66$   & $1.523$ & $152$ & $1.05$ & $ 3.38 $ \\       

  09 Dic & 17 & $1182.66$  & $1.534$ & $219$ & $2.60$ & $ 2.40 $ \\

  09 Dic & 18 & $1186.66$  & $1.400$ & $58$  & $0.39$ & $ 2.06 $ \\

  09 Dic & 19 & $1191.66$  & $1.488$ & $94$  & $1.24$ & $ 2.84 $ \\

  10 Jan & 21 & $1205.66$  & $1.510$ & $120$ & $1.04$ & $ 2.23 $ \\

  10 Jan & 22 & $1209.66$  & $1.493$ & $98$  & $0.95$ & $ 4.76 $ \\       

  10 Mar & 23 & $1274.66$  & $1.430$ & $66$  & $0.68$ & $ 2.99 $ \\       

  11 Jun & 24 & $1739.66$  & $1.460$ & $76$  & $0.74$ & $ 3.16 $ \\              

  11 Jul & 25 & $1745.66$  & $1.527$ & $169$ & $0.81$ & $ 8.09 $ \\              

  11 Jul & 26 & $1766.66$  & $1.469$ & $81$  & $0.36$ & $ 7.13 $ \\              

  11 Aug & 28 & $1783.66$  & $1.380$ & $55$  & $0.41$ & $ 2.40 $ \\              

  11 Oct & 29 & $1848.66$  & $1.460$ & $76$  & $0.67$ & $ 3.30 $ \\              

  11 Oct & 30 & $1853.66$  & $1.541$ & $383$ & $1.32$ & $ 24.52 $ \\              

  11 Nov & 31 & $1867.66$  & $1.522$ & $149$ & $0.57$ & $ 16.83 $ \\              

  11 Nov & 33 & $1875.66$  & $1.531$ & $193$ & $0.88$ & $ 6.37 $ \\              

  12 Feb & 36 & $1972.66$  & $1.220$ & $39$  & $0.66$ & $ 3.55 $ \\              

  12 Mar & 37 & $1982.66$  & $1.350$ & $50$  & $2.03$ & $ 7.48 $ \\              
                           
  \hline
                   
 \end{tabular}
  \caption{Different physical quantities obtained for the outbursts
modelled in this work.  The background Lorentz factor of the bulk velocity
of the flow was assumed to be \( 18 \). The first three columns from
left to right are the date, numeric identification of the outburst
(ID \#) and the date corresponding to the maximum luminosity for a
particular outburst. Columns four and six are the obtained values for the
parameters \( \eta^2\) (measured in units of the speed of light \( c \))
and the inverse frequency \( \omega^{-1} \) relevant to the particular
variational model of equation~\eqref{eq01}.    Column five corresponds to
the maximum upper limit of the Lorentz factor of the flow for each
particular outburst. The minimum Lorentz factor of the flow for
all outbursts is \( \sim 12 - 13 \).  Column seven represents the mass
injection rate \( \dot{m} \) of the flow at the base of the jet. The values of 
all inferred parameters are accurate to within 10\%. }

 \label{table1}
 \end{footnotesize} 
 \end{center}
\end{table}
%

%

%
%

\section{Discussion}
\label{discussion}

  We have modelled the LC of Blazar PKS 1510-089 for almost 4 years
using the hydrodynamical model of M09.  The modelling was performed
by assuming a periodic velocity injection mechanism at the base of the
relativistic jet that leads to the formation of a working surface and is
capable of loosing energy as it travels along the jet.  As explained in
section~\ref{model}, the model by M09 was constructed to deal with LCs of
lGRB. However, the Blazar PKS 1510-089 has many physical characteristics
to be considered a geometrical large scaled version of a lGRB since
it has a highly relativistic jet that points towards the observer.
The results presented in Table~\ref{table1} show high upper limits for
the bulk Lorentz factors achieved with oscillations of the flow, that
reach values as large as \( \lesssim 380 \) for one particular event.
These inferred huge Lorentz factors in the bulk velocity oscillation of
this Blazar show another close similarity with lGRB's.


 The range of parameters as presented in Table~\ref{table1}, i.e.  
\( \dot{m} \sim \left( 2 - 25 \right) \times 10^{-3}  \textrm{M}_\odot
\textrm{yr}^{-1} \), \( \omega^{-1} \sim \left( 0.3 - 2.6 \right)
\times 10^{3} \, \textrm{s} \)  and variations of the Lorentz factor \(
\Gamma \sim 10 - 380 \), denote a scaling between the lGRB counterparts
found in M09 for which \( \dot{m} \sim 10^{-1}-10^{-2} \textrm{M}_\odot
\textrm{s}^{-1}\), \(\omega^{-1} \sim\ 10 \textrm{s}\) and \(\Gamma \sim
50 - 500\).  Note that the maximum and minimum values of the Lorentz
factor for a particular outburst take into account the observational
errors of the LC.  The real value lies in between those calculated ranges.
The inferred high relativistic Lorentz factors associated to the motion of
the bulk velocity of the flow inside the jet of PKS~1510-089 makes it an
ideal candidate for the application of the hydrodynamical model of M09.
This is why that physical model can be applied naturally to lGRB and
in this particular case to the extreme relativistic motion of the jet
in the Blazar PKS~1510-089.  The energy released in each outburst can
be calculated by taking the integral of the luminosity with respect to
time, which occurs typically over periods of a few days.  The value of
this released energy  is \( \sim 10^{39} \) -- \( 10^{40} \)~\text{J},
which shows the tremendous energy released by each individual outburst.
This energy is to be compared with the energy released in about \(
10~\text{s} \) by a lGRB which is \( \sim 10^{44}~\text{J} \).

The most energetic burst, labelled 30, inyected at the base of the jet
a total mass \( m = \dot{m} \Delta t \sim 10^{-3} \, M_\odot\)
while the burst lasted \( \Delta t \sim 15 \text{days} \). Analysis of
all bursts shows that the ejected mass interval is \( 10^{-5} M_\odot
\lesssim m \lesssim  10^{-3} M_\odot \), for a time duration range \(
4 \text{days} \lesssim \Delta t \lesssim 30 \text{days} \).

The variations of the
injected flow at the base of the jet cause the formation of working
surfaces that produce bursts of \(\gamma\)-rays in the
structure of the jet.  The physical mechanism producing the oscillations
of the input flow, which allows fast fluid to overtake the slow one,
leading to the formation of working surfaces, is beyond the scope of
this letter. However, steady flow deviations and oscillations in such
complicated phenomena are expected since the accretion-ejection mechanism
associated to a particular object is not necessarily expected to be of
constant velocity and mass accretion-ejection rates.

 It is important to note that the assumption of seeing a Blazar as a
scaled version of a lGRB is not new.  In an early attempt for finding
a unified model of jet and central-engine power, \citet{mirabel02} made
this identification.  The more relativistic a Blazar jet is, the more it
will resemble a lGRB.  The idea of having a unified physical model for
all types of astrophysical jets was first suggested by the pioneering
works for the astrophysical scaling laws of black holes by \citet{sams96}
and \citet{rees98}.  The work presented in this letter further strengths
arguments about a unified picture of all astrophysical relativistic jets.

  PKS 1510-089 resulted to be an ideal target to test the model by
M09 since it closely resembles a lGRB in some of its
outbursts.  Future tests of the model have to be done with a wide variety
of Light Curves from a large collection of Blazars and micro-quasars.
\section{ACKNOWLEDGMENTS}
We thank the anonymous referee for his valuable comments who helped us to
produce a much improved version of our letter.  This work was supported by
three DGAPA-UNAM grants (PAPIIT IN116210-3, IN116211-3, IN111513-3). JIC
acknowledges support given by IAUNAM as a visiting researcher.  JIC,
YUC, EB, SM, DH and MS thank support granted by CONACyT: 50102, 210965,
13654, 26344, 8366, 177304.   The authors acknowledge the use of the
\textit{FERMI-LAT} publicly available data as well as the public data
reduction software.

%
%
\label{lastpage}
\bibliographystyle{mn2e-extra}
\bibliography{letterpks}

\end{document}